\documentclass[aps,prd,10pt,twocolumn,nofootinbib,floatfix]{revtex4}
\usepackage[title]{appendix}
\usepackage{epsfig}
\usepackage{bm}
\usepackage{latexsym}
\usepackage{natbib}
\usepackage{url}
\usepackage{dcolumn}
\usepackage{color}
\usepackage{amsfonts,amssymb,amsmath}
\usepackage{graphicx,epsfig}
\usepackage{psfrag}
\usepackage{subfigure}
\usepackage{hyperref}
\hypersetup{colorlinks=true}
\usepackage{mathtools}
\usepackage{enumitem}
\usepackage{float}
\usepackage{mathrsfs}
\usepackage{xcolor}
\hypersetup{colorlinks,  citecolor=red,  linkcolor=red,  urlcolor=blue}
\usepackage{comment}
\usepackage[normalem]{ulem}
\usepackage{cleveref}
\usepackage{nameref}
\usepackage{aas_macros}

\begin{document}

\title{Probing self-interacting ultrahigh-energy neutrinos with the cosmic 21-cm signal}

\author{
Mansi Dhuria\footnote{Mansi.dhuria@sot.pdpu.ac.in},
Bishnu Gupta Teli\footnote{bishnu.tbsc20@sls.pdpu.ac.in}
}

\affiliation{
Department of Physics, School of Energy Technology, Pandit Deendayal Energy University (PDEU), Gandhinagar-382426, Gujarat, India
}

\begin{abstract}
\noindent 

In this study, we investigate the constraints on secret self-interactions of neutrinos by examining the impact of radiative scattering of ultrahigh-energy neutrinos. These neutrinos are produced from the decay of superheavy dark matter and interact with the cosmic neutrino background. We explore how these interactions influence the 21-cm  hydrogen signal during the cosmic dark ages and cosmic dawn, periods relatively free from astrophysical uncertainties, providing a clearer signal for studying nonstandard neutrino interactions. By analyzing the global brightness temperature measurements, we constrain the scattering cross section of ultrahigh-energy self-interacting neutrinos, determining the coupling constant $g$ to be within $\sim 10^{-4}$ to $\sim 10^{-3}$ for neutrino energies in the PeV to EeV range. Interestingly, these constraints are more competitive than those from existing astrophysical and collider experiments.
As future 21-cm experiments focus on measuring brightness temperature across a wide range of redshifts from the cosmic dark ages to reionization, using the epoch of 21-cm to probe neutrino properties could provide crucial insights into dark matter and neutrino physics.

\end{abstract}

\maketitle

\section{Introduction}
\label{sec:Introduction}
Despite the significant progress made in observing our Universe through various methods such as galaxy surveys, cosmic microwave background (CMB) based measurements, and recent gravitational interferometers, etc., large portions of our Universe remain unexplored, particularly in the redshift range between $z \sim 1100$ and $z \sim 6$ due to the faintness of early Universe sources. Remarkably, recent years have revealed another promising avenue of exploration spanning from shortly after the epoch of recombination at redshift $z \sim 1100$ to the formation of the initial significant population of luminous objects around redshift $z \sim 30$, up to the reionization of the Universe at redshift $z \sim 6$~\cite{liu2022snowmass2021cosmicfrontierwhite, Bera:2022vhw}. During this epoch, the Universe was primarily governed by neutral hydrogen until the emergence of the first stars and galaxies. Consequently, much of the investigation remains centered on observing the 21-cm signal emitted by neutral hydrogen, originating from its hyperfine transition~\cite {Pritchard_2012}. The significant interest in such models has been sparked by the potential detection of a robust 21-cm signal by the EDGES experiment~\cite{Bowman_2018}. The signal was significantly stronger than the maximal absorption signal possible within standard cosmology, hinting towards nonstandard dynamics~\cite{Barkana:2018lgd}. Although the SARAS experiment~\cite{Singh:2021mxo} has contested it at a 95\% significance level, more results are needed to rule out the EDGES claim. Other than that, several other ongoing and future experiments such as LEDA~\cite{Price_2018}, and REACH~\cite{deLeraAcedo:2022kiu, Saxena:2024rhu}, focus on detecting the global 21-cm signal across a wide range of redshift, while various radio interferometric telescopes such as MWA~\cite{tingay2013murchison}, GMRT~\cite{1991CSci...60...95S}, LOFAR~\cite{LOFAR:2013jil}, HERA~\cite{DeBoer:2016tnn}, and SKA~\cite{Koopmans:2015sua} are dedicated to probing the spatial fluctuations in the 21-cm hydrogen signal during the period of cosmic dawn and reionization. Given these ongoing/upcoming facilities, we must have a thorough understanding of the cosmological 21-cm hydrogen signal expected in consistent models of cosmology.

In the past few years, it has been well established that the observed 21-cm signal is greatly influenced by interactions between dark matter (DM) and baryons. Recent studies have delved into examining the influence of various DM candidates and their interactions on the 21-cm observables such as cooling of hydrogen gas due to elastic scattering with DM~\cite{Natwariya_2022, Tashiro:2014tsa, Munoz:2015bca, Munoz:2018pzp, Barkana:2018lgd, Barkana:2018qrx, Berlin:2018sjs, Dhuria:2018kzb, Dhuria:2021lqs, Dey:2022ini, Barkana:2022hko, Mondal:2023bxb}, heating of hydrogen due to decay or annihilation of DM~\cite{Evoli:2014pva, Lopez-Honorez:2016sur, Liu:2018uzy, DAmico:2018sxd, Cheung:2018vww, Mitridate:2018iag, Clark:2018ghm, Mittal:2021egv, Halder:2021uoa, Acharya:2023ygd, Facchinetti:2023slb, basu2023addressingselfinteractionelderdark, cang2023signaturesinhomogeneousdarkmatter, Shao:2023agv, sun2023inhomogeneousenergyinjection21cm}, modifications to the Rayleigh-Jeans tail due to the resonant conversion of DM to CMB photons, etc~\cite{Pospelov:2018kdh, Bondarenko:2020moh, Caputo:2020avy, Caputo:2022keo}.

In this work, we study the impact of secret self-interactions of UHE neutrinos emitted from DM decay on the 21-cm brightness temperature. 
Neutrinos are known for their extremely weak interactions via the weak force in the standard model (SM), thus making it easy to travel through Earth. The recent cosmological and astrophysical observations have highlighted the significant role of neutrinos in multimessenger astronomy. 
Particularly, the UHE neutrino flux from DM decay/annihilation or other astrophysical sources as probed by numerous current-generation neutrino experiments such as IceCube has been able to set competitive limits on the DM lifetime and  DM annihilation cross section for very heavy DM candidates~\cite{IceCube:2018tkk, IceCube:2023orq, IceCube:2023gku}. In the last few years, there has also been increasing discussion about the potential for neutrinos to interact quite strongly with themselves via a new scalar/vector mediator, termed neutrino self-interaction ($\nu{\rm SI}$)~\cite{Berryman:2022hds, Belotsky:2001fb, Berkov:1988sd, Berkov:1987pz}. This phenomenon has implications such as addressing the Hubble tension \cite{Oldengott_2017, Forastieri_2019, Das:2020xke, Choudhury_2021,Das:2023npl}, supporting KeV sterile neutrino as a viable DM candidate \cite{de_Gouv_a_2020,Kelly:2020pcy,Benso:2021hhh, Dhuria_2023, Chen_2022,An:2023mkf}, and influencing supernova neutrino emission~\cite{Kolb:1987qy, Chen:2022kal, bhattacharya2023gravitationalmemorysignalneutrino}. Thus, the existence of $\nu{\rm SI}$ naturally suggests physics beyond the SM, offering opportunities to explore its implications for various astrophysical and cosmological phenomena such as \cite{Dhuria:2023itq, hyde2023constraintsneutrinoselfinteractionsicecube, RoyChoudhury:2022rva, Wu:2023twu, Bostan:2023ped}.

 The investigation into the secret self-interaction of UHE neutrinos is also conducted in literature through the scattering of UHE astrophysical neutrinos with cosmic neutrino background (C$\nu$B) neutrinos ~\cite{Hooper:2007jr, Ioka:2014kca, Ng:2014pca, Bustamante:2020mep, Esteban_2021}. The interaction between UHE neutrinos and C$\nu$B neutrinos results in distinctive dips and bumps in the astrophysical spectrum. By comparing this spectrum with the current data from IceCube, constraints on the self-interacting coupling of $\tau$-neutrinos have been established~\cite{Esteban_2021}. Although the existing IceCube data has been able to probe very high values of self-interacting coupling, it has been shown that the upcoming IceCube data should be able to probe even moderately small values of couplings~\cite{Esteban_2021}. In fact, the bounds on the self-interacting couplings obtained from the same are much stronger than those from other cosmological and collider probes~\cite{Blinov:2019gcj, Lyu:2020lps}. While neutrino experiments can constrain such interactions or the annihilation/scattering cross section of DM, distinguishing UHE neutrinos from DM decay from those emitted by astrophysical sources remains a significant challenge. The effectiveness of neutrino telescopes in investigating heavy DM relies heavily on the diffuse flux spectrum of UHE neutrinos. These UHE neutrinos act as a background to the flux spectrum from DM searches, potentially masking the subtle signs of DM decay~\cite{Fiorillo:2023clw}. Hence, this remains a considerable challenge. Recently, there has been an attempt to investigate interactions of superheavy DM by searching for radio emissions resulting from the interaction of UHE neutrinos with the lunar regolith~\cite{das2024probingsuperheavydarkmatter}. This method can potentially explore energy levels beyond  $10^{12}$ GeV, which are beyond the reach of astrophysical accelerators.

In our work, we aim to constrain secret self-neutrino interactions of UHE neutrinos emitted specifically through the decay of superheavy DM by studying its impact on the 21-cm signal in the period from dark ages to cosmic dawn, which is almost free from the astrophysical background. In a toy physics model beyond the SM, the self-interaction between neutrinos can be introduced by involving a new scalar or vector boson that interacts with a pair of neutrinos and their leptonic partners. For simplicity, we consider only a scalar boson in this work. In addition to elastic scattering between neutrinos, new interactions can also lead to the production of photons due to the radiative (one-loop) scattering of UHE neutrinos with cosmic neutrinos mediated by leptonic partners and the scalar boson. This process can heat the intergalactic medium (IGM), thereby affecting the 21-cm brightness temperature. By considering the number density of UHE neutrinos allowed by the present-day relic abundance of DM and their interaction with the cosmic neutrino background in the redshift ranging from the dark ages to cosmic dawn, we study the impact of such heating on the 21-cm absorption spectrum. As a result, we obtain constraints on the allowed parameter space of self-interacting neutrinos and find that these constraints are much more competitive than those from other astrophysical and laboratory probes. 
As many upcoming radio experiments aim to detect the 21-cm brightness temperature and the 21-cm power spectrum more precisely across a wide redshift range, our analysis will be valuable in highlighting the potential signatures of nonstandard neutrino interactions using the 21-cm absorption signal in the future.

The plan for the rest of the paper is as follows: In Sec. \ref{sec:21-cm Cosmology}, we begin with a brief overview of the 21-cm cosmology. In Sec. \ref{sec:Self-interacting UHE neutrinos from decay of heavy DM}, we discuss a basic toy model of particle physics that involves the interaction of a scalar boson with a pair of neutrinos and their leptonic partners. We also explain the possibility of producing photons through the radiative scattering of UHE neutrinos with C$\nu$B neutrinos. In Sec. \ref{sec:Effect of DM induced heating on 21-cm signal}, we discuss the general steps in calculating heating induced by various processes on the gas temperature and the free electron fraction. In Sec. \ref{sec:Energy Deposition Rate due to self-scattering of UHE neutrinos}, we specifically calculate the energy injection rate due to
the self-scattering of UHE neutrinos emitted from superheavy DM with C$\nu$B background. In Sec. \ref{sec:Results and Discussion}, we present our findings on the impact of specific scattering cross section values on the 21-cm absorption signal. Additionally, we derive constraints on the parameter space for self-interacting neutrino strength, considering the mass of the scalar mediator and various levels of UHE neutrino emitted from superheavy DM. In Sec. \ref{sec:Concluding Remarks}, we summarize our results with conclusions and suggest possible future directions. There is one Appendix.

\section{21-cm Cosmology}
\label{sec:21-cm Cosmology}
We start with a concise overview of the fundamental aspects of 21-cm cosmology~\cite {Pritchard_2012}.  The CMB photons we observe today have traveled through the Universe since a redshift of around 1100, passing through cold neutral hydrogen clouds. During their journey, 21-cm wavelength photons were absorbed and emitted via hydrogen's hyperfine transitions. This process causes a deviation in the CMB spectrum, typically quantified by the differential 21-cm brightness temperature.
The global 21-cm differential brightness temperature is defined as~\cite{Pritchard_2012}
\begin{equation}
    T_{21}=27x_{HI}\left[\frac{0.15}{\Omega_m}\frac{1+z}{10}\right]^{1/2}\left(\frac{\Omega_bh}{0.023}\right)\left(1-\frac{T_\gamma}{T_s}\right)\ mK
\label{eq:T21}
\end{equation}
where $x_{HI}=n_{HI}/n_H$ is the fraction of neutral hydrogen in the Universe, $n_{HI}$ and $n_H$ are the number densities of neutral hydrogen, and total hydrogen (ionized+neutral), respectively. $T_\gamma$  corresponds to the temperature of the surrounding bath of photons, typically fixed by the CMB temperature, so that $T_\gamma=T_{\rm CMB}$. Here, $\Omega_b\approx0.044$ is relic abundance of baryonic matter, and $\Omega_m\approx0.26$ is relic abundance of total matter. The parameter $h$ is the reduced Hubble constant and its value is $h=\frac{H_0}{100\ km/s/Mpc}=0.74$.
The relative populations between the two hyperfine levels, triplet and single state of a neutral hydrogen atom, can be determined by the spin temperature parameter  $T_s$, defined as $n_1/n_0=3e^{-T_*/T_s}$. The splitting between the singlet and triplet state is denoted by $\frac{\Delta E}{k_B}=T_*=0.068K$. The spin temperature is generally influenced by three factors~\cite{Short:2019twc}: (i) absorption/emission resulting from Compton scattering with the surrounding CMB photons, (ii) collisional coupling between hydrogen molecules, which is more significant at high redshifts, and (iii) resonant scattering with the Lyman-$\alpha$  (Ly$\alpha$) photons, generally known as the Wouthusysen-field effect~\cite{Hirata_2006}. The evolution of the spin temperature is given as~\cite{1958PIRE...46..240F}
\begin{equation}\label{spintemperature}
    T_s^{-1}=\frac{T_{\gamma}^{-1}+x_{c}T_k^{-1}+x_\alpha T_c^{-1}}{1+x_c+x_{\alpha}}
\end{equation}
where $T_k$ is the kinetic gas temperature and $T_c$ is the color temperature of the Ly$\alpha$ photons at the Ly$\alpha$ frequency. In most relevant scenarios, $T_c \approx T_K$ because the optical depth to Ly$\alpha$ scattering is usually quite high. This results in numerous scatterings of Ly$\alpha$ photons, which align the radiation field and the gas near the line center frequency, achieving local equilibrium \cite{1959ApJ...129..551F, Pritchard_2012}. The parameters $x_c$ and $x_\alpha$ are the coupling coefficients due to atomic collisions and scattering of Ly$\alpha$ photons, respectively.

As the collisional coupling is mainly induced by collisions between hydrogen atoms with other hydrogen atoms, free electrons,
and free protons, the total coupling coefficient will be given by
\begin{align}
    x_c &= x_c^{HH} + x_c^{eH} + x_c^{pH},
    \end{align}
    The collision coupling coefficient \cite{Furlanetto_2006} for a particular channel is
\begin{equation}
    x_c^i=\frac{n_i\kappa_{10}^i}{A_{10}}\frac{T_*}{T_{\gamma}}
\end{equation}
where $\kappa^i_{10}$ denotes the rate coefficient for spin deexcitation in collisions in that particular channel (with units of $m^3s^{-1}$). With this, the coupling coefficient $x_c$ turns out to be
\begin{align}
    x_c  
    &= \frac{T_*}{A_{10}T_\gamma}\left[\kappa^{HH}_{10}(T_k)n_H + \kappa^{eH}_{10}(T_k)n_e + \kappa^{pH}_{10}(T_k)n_p\right]\label{x_c(expanded)}
\end{align}
where $\kappa^{HH}_{10}$ is the scattering rate between hydrogen atoms~\cite{Furlanetto_2006} , $\kappa^{eH}_{10}$ is the scattering rate between electrons and hydrogen atoms~\cite{Furlanetto_2007e}, and $\kappa^{pH}_{10}$ is the scattering rate between protons and hydrogen atoms~\cite{Furlanetto_2007p}.

In the beginning, at high redshifts, the collisional coupling is dominating. But at the lower redshifts, the collision coupling becomes subdominant as the number density of hydrogen gas decreases. However, after the formation of the first star, resonant scattering of Ly$\alpha$ photons provides a new way of coupling. This is famously called the Woulthuysen-field mechanism \cite{1952AJ.....57R..31W}. The physics of this mechanism is much more subtle than this description. For convenience, we will use the seminumerical model to calculate the coupling constant ($x_\alpha$). We consider tan$h$ parametrization model given in Refs.~\cite{PhysRevD.98.103529, Pritchard_2012, 2016MNRAS.455.3829H} in order to calculate coefficient $x_\alpha$ given by\footnote{The tan$h$ method for modeling the global 21-cm signal has been widely adopted as one of the computationally efficient parametrization model. While there are other intermediate approaches—such as the ``turning points'' parametrization (outlined in~\cite{Pritchard_2012, 2016MNRAS.455.3829H}) or the Gaussian-based absorption feature (discussed in \cite{Bernardi:2014caa}), it has been discussed in \cite{Mirocha:2015jra} that these approaches fail to accurately capture the detailed shape of physically motivated models. Moreover,  Ref. \cite{2016MNRAS.455.3829H} explicitly demonstrates that the tanh parameterization fits well with 21-cm signals computed using the accelerated reionization era simulations code.}
\begin{equation}
    x_\alpha(z)\equiv 2A_\alpha(z)/(1+z)
\end{equation}
where
\begin{align*}
    A_\alpha(z)=A_\alpha\left(1+{\rm tan}h\left(\frac{z_{\alpha0}-z}{\Delta z_\alpha}\right)\right)
\end{align*}
 As suggested in \cite{PhysRevD.98.103529}, we use following set of fiducial values to calculate $x_\alpha$:
$$\{A_\alpha,z_{\alpha 0},\Delta z\}=\{100,17,2\}$$

Using the expressions of the coupling coefficients $x_\alpha$ and $x_c$, the spin temperature $T_s$ can be calculated by using Eq.~(\ref{spintemperature}) and consequently brightness temperature $T_{21}$ can be calculated by using Eq.~(\ref{eq:T21}) in the relevant range of redshift. In standard cosmology, one expects two absorption signals with the first shallow absorption minima near 20 MHz (with $z \sim
70$) and the other deeper minima at higher frequencies between $50$$-$$110$ MHz (with $z \sim 12$$-$$27$) in the
global cosmological 21-cm signal, which are signatures of collisional gas dynamics in the
cosmic dark ages and Ly$\alpha$ photons from the first stars at cosmic dawn, respectively. 

\section{Self-interacting UHE neutrinos from decay of dark matter}\label{sec:Self-interacting UHE neutrinos from decay of heavy DM}
We consider a scenario in which a superheavy DM particle with a mass $m_{\rm DM} \geq {\rm PeV}$ predominantly decays into UHE neutrinos. The number of neutrinos produced depends on the specific decay channels of the DM particle. If the heavy DM primarily undergoes two-body decay into a pair of neutrinos, i.e., ${\rm DM} \rightarrow {\nu} {\bar \nu}$, this will lead to a neutrino flux with $E_{\nu_h} \approx m_{\rm DM}c^2$. As mentioned in Sec. \ref{sec:Introduction}, the UHE neutrino flux from DM as well as astrophysical sources is being probed by numerous current-generation neutrino experiments~\cite{Esteban_2021}. These experiments consider the scattering of high-energy neutrinos with the C$\nu$B en route to Earth, which redistributes their energies. We will consider the UHE neutrinos emitted specifically from the decay of DM and study the impact of the scattering of the same with the C$\nu$B on the 21-cm signal during the period from dark ages to cosmic dawn.

 \begin{figure*}[htbp]
  \centering
  \subfigure[]{\includegraphics{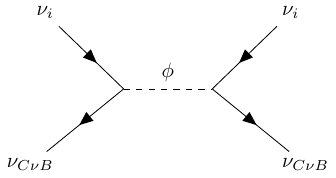}}
  \hskip 1.0in \subfigure[]{\includegraphics{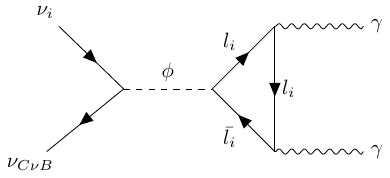}}
  \caption{Feynmann diagram showing (a) tree-level scattering of UHE neutrinos with C$\nu$B neutrinos, and  (b) one-loop level scattering of UHE neutrinos with C$\nu$B neutrinos into UHE photons and CMB background photons.}
  \label{fig-feynmann}
\end{figure*}

In the minimal model of neutrino self-interaction, we consider a model in which the
real singlet scalar at low energies couples both to  
neutrinos as well as leptonic partners. The interaction couplings
are given as:
$$L \supset g_{\nu_{i}}\phi\nu_{i}\nu_{i} + g_{l_i}\phi\bar{l_i}l_i$$
where $i = e, \mu, \tau$
represent three different flavors of neutrinos. In this formulation, we assume the Majorana neutrinos and use Weyl notation to denote the neutrino coupling to scalar bosons and Dirac notation to denote leptons coupling to the scalar boson. Similarly, the interaction can also be mediated through the new $Z$ boson. In the context of a scalar boson, the coupling parameters $g_{l_i}$
  and $g_{\nu_i}$ may either be identical or distinct for a particular generation, depending on the specific particle physics model.
In most general scenarios, the coupling $g_{\nu_i}$ is determined by the mechanism responsible for neutrino mass generation, while  $g_{l_i}$ arises from nonrenormalizable interactions involving the Higgs field and a new scalar field. 
Since the leptonic coupling depends on the mass of the leptons, we cannot expect the coupling $g_{l_i}$ to be same for all generations of leptons. Thus, for the sake of simplicity in our toy model, we assume $g_{l_i} = g_{\nu_{i}} = g_i$ for a particular generation.

In the presence of aforementioned interactions, the tree-level $s$-channel scattering of neutrinos can induce self-scattering of neutrinos, while the one-loop scattering mediated through leptons and a new scalar can produce photons. The tree-level and one-loop level Feynman diagram for this process is given in Fig.~\ref{fig-feynmann}, respectively. The emission of gamma rays produced by radiative scattering can heat the IGM, which, in turn, can affect the 21-cm global signal during the cosmic dawn and the dark ages. The cross section for the one-loop process, shown in the Feynman diagram is given by \cite{Dhuria:2023itq}
\begin{equation}\label{scatteringcrosssection}
    \sigma = \frac{81\alpha^2 s}{4\pi^3}\frac{g^4_i}{(s-m_\phi^2)^2+m_\phi^2\Gamma_\phi^2}\times\left|1+ Q_i^2m_i^2 C_0^\gamma\right|^2
\end{equation}
where $C_0^\gamma$ is known as the scalar Passarino-Veltaman function and is given by,
\begin{equation}\label{Passarino-Veltman Function}
    C_0^\gamma(s,m_i)=\frac{1}{2s}\ln^2\left(\frac{\sqrt{1-4m_i^2/s}-1}{\sqrt{1-4m_i^2/s}+1}\right)
\end{equation}
In the above Eqs.~(\ref{scatteringcrosssection}) and (\ref{Passarino-Veltman Function}), $g_i$ stands for the self-interacting coupling for a particular flavor of neutrinos, $m_i$ and $Q_i$ represent the mass and electromagnetic charge of the corresponding leptonic partner, respectively, $s=2m_{\nu_i}E_{\nu_h}$ stands for the center of mass Mandelstam variable, $m_\phi$ is the mass of the new scalar mediator and $\Gamma_\phi=g^2_i m_\phi/4\pi$ is the neutrino decay width. Here, $m_{\nu_i}$ is the mass of the active neutrino of a particular flavor. Depending on the values of $E_{\nu_h}$ and $m_\phi$, this cross section can reach a resonance when $E_{\nu_h}\approx m_\phi^2/2m_{\nu_i}$. 

Using the expression for the cross section given in Eq.~(\ref{scatteringcrosssection}), we will now study the effect of self-interactions of UHE neutrinos on the evolution of 21-cm brightness temperature. 

\section{Effect of heating on the 21-cm signal}\label{sec:Effect of DM induced heating on 21-cm signal}
In this section, we will first discuss the general effect of heating induced by new physics and its consequent impact on the cosmic 21-cm absorption signal. The energy injection from such heating can alter the temperature of hydrogen gas during the cosmic dark ages and cosmic dawn, affecting the absorption of a 21-cm signal. Before quantifying the energy injection resulting from the scattering of UHE neutrinos with the cosmic neutrino background, we will outline the steps involved in calculating the evolution of the gas temperature ($T_k$) and ionization fraction ($x_e$) due to standard cosmological effects, as well as the additional effects due to heating.

To calculate the brightness temperature given by Eq.~(\ref{eq:T21}), one needs to calculate the evolution of the fractional neutral hydrogen ($x_{HI}$), CMB temperature ($T_{\rm CMB}$) and spin temperature ($T_s$) as a function of redshift. The parameter $x_{HI}$ is related to the fraction of ionized hydrogen ($x_e$) as $x_{HI}=1-x_e$. The CMB temperature can be calculated as $T_{\rm CMB}=T_{\rm CMB,0}(1+z)$, where $T_{\rm CMB,0}=2.7$K is the CMB temperature today. From Eq.~(\ref{spintemperature}), we can see that in addition to $T_{\rm CMB}$, the spin temperature also depends on $x_c$, $x_\alpha$, $T_k$ and $T_c$. Explanation to calculate $x_c$, $x_\alpha$ and $T_c$ is given in Sec. \$\ref{sec:21-cm Cosmology}. Thus, the evolution of brightness temperature essentially depends on the history of gas temperature ($T_k$) and ionization fraction ($x_e$). To calculate the same, we follow a standard Peebles recombination framework \cite{1968ApJ...153....1P}, further refined in subsequent studies \cite{Seager_1999, 10.1111/j.1365-2966.2009.15957.x, Ali_Ha_moud_2011}. This framework involves solving two coupled ordinary differential equations for the evolution of gas temperature and ionization fraction. 
\subsection{Evolution of gas temperature}
\label{subsec:Evolution of gas temperature}
The evolution of the kinetic gas temperature with redshift follows~\cite{PhysRevLett.121.011103, PhysRevD.98.023501}:
\begin{equation}\label{odeTk1}
    \frac{dT_k}{dz}=\frac{2 T_k}{1+z}+\frac{\Gamma_C}{(1+z)H}(T_k-T_{CMB})
\end{equation}
In Eq.~(\ref{odeTk1}), the first term shows the effect of cosmological expansion on the gas temperature.
The second term represents the heating due to the Compton scattering between hydrogen gas and CMB photons. Here, 
$\Gamma_c$
  denotes the Compton scattering rate, defined as
\begin{equation}\label{Comptonscatteringrate}
    \Gamma_C=\frac{8\sigma_Ta_rT^4_{\rm CMB}x_e}{3(1+f_{He}+xe)m_ec}
\end{equation}
where $\sigma_T$, $a_r$ and $m_e$ are the Thomson scattering cross section, Stephan-Boltzmann radiation constant and mass of an electron, respectively and $f_{He}=n_{He}/n_H$ is the helium fraction. It has also been shown in Ref.~\cite{PhysRevD.98.103513} that Lyman-$\alpha$ photons facilitate energy transfer between CMB photons and the thermal motions of hydrogen atoms. In scenarios lacking x-ray heating, this newly identified mechanism significantly modulates the temperature of adiabatically cooling gas by approximately $10\%$ at $z\approx17$. To include this effect, Eq.~(\ref{odeTk1}) gets modified as,
\begin{align}\label{odeTk2}
    \frac{dT_k}{dz}=\frac{dT_k}{dz}\bigg|_{eq. (\ref{odeTk1})}-\frac{\Gamma_R}{H(z)(1+z)}\left(\frac{T_\gamma}{T_s}-1\right)T_*
\end{align}
where $\Gamma_R$ is the heating rate due to the transfer of energy from CMB photons to the thermal motion of hydrogen gas and is given by
\begin{align}
    \Gamma_R=\frac{x_{HI}x_{\rm CMB}}{2(1+f_{He}+x_e)}A_{10}
\end{align}
where $A_{10}=2.86\times10^{-15}\ s^{-1}$ is the Einstein coefficient for spontaneous emission from the triplet state to singlet state, 
and
\begin{equation}
    x_{CMB}=\frac{1}{\tau_{21}}(1-e^{-\tau_{21}})
\end{equation}
where
$\tau_{21}$ is optical depth given by
\begin{equation}
    \tau_{21}=8.1\times10^{-2}x_{HI}\left(\frac{1+z}{20}\right)^{1.5}\frac{10\ K}{T_s}.
\end{equation}
Finally to include additional injection of energetic particles due to certain BSM processes such as DM decay/annihilation, decay of primordial black holes, or the scattering of energetic particles, etc., Eq.~(\ref{odeTk2}) can be modified as follows:
\begin{align}\label{odeTk}
    \frac{dT_k}{dz}&=\frac{dT_k}{dz}\bigg|_{eq. (\ref{odeTk2})}\\
    &-\frac{2}{H(z)(1+z)3k_Bn_H(z)(1+f_{He}+x_e)}\frac{dE}{dV dt}\bigg |_{dep,h}\notag
\end{align}
where $k_B$ is the Boltzmann constant. The last term in Eq.~(\ref{odeTk}) corresponds to the energy deposition into the IGM. Each channel of energy deposition is represented by the subscripts $c={i,\alpha,h}$, corresponding to ionization, excitation, and heating, respectively. It should be noted that not all of the energy injected from DM interaction is fully deposited in the medium. The quantity of energy deposited in the medium heavily depends upon various DM interaction channels. The energy deposition rate \cite{Mitridate_2018, Slatyer_2013, Finkbeiner_2012}, in general form is given in terms of energy injection rate as,
\begin{equation}\label{energydeposited}
    \frac{dE}{dV dt}\bigg|_{dep,c} = f_c(z)\frac{dE}{dV dt}\bigg|_{inj}
\end{equation}
where $f_c(z)$ is a dimensionless factor representing efficiency, the amount of deposited energy in the medium in the three different channels. In this work, we assume that a fraction $f_{\rm eff}$ of the energy produced by various processes such as DM and heavy particle decay/annihilation, scattering, etc. at certain redshift is instantaneously transferred to the plasma, using a simplified approach called the ``SSCK'' approximation \cite{Chen_2004, PhysRevLett.121.011103}.
\begin{equation*}
    f_{i}=f_{\alpha}\approx f_{eff}\frac{1-x_e}{3},\hspace{0.5cm}
    f_{h}=f_{eff}\frac{1+2x_e}{3}
\end{equation*}
For all our analysis, we will be using $f_{\rm eff}\approx0.1$, as discussed in~\cite{Short_2020}.
 
\subsection{Evolution of free electron fraction}
The evolution of the ionization fraction/free electron fraction ($x_e$) with redshift ($z$) is given by \cite{Chen_2004},
\begin{equation}\label{odexe}
    \frac{dx_e}{dz}=\frac{1}{(1+z)H(z)}[R_s(z)-I_s(z)-I_X(z)]
\end{equation}
where $R_s$ and $I_s$ are the standard recombination rate (from ionized gas to neutral gas) and standard ionization rate (from neutral gas to ionized gas). The details of these parameters are given in the Appendix.

The last term in Eq.~(\ref{odexe}), $I_X$, can be written as $I_X=I_{X_{i}}+I_{X_{\alpha}}$. It represents an ionization rate due to the additional injection of energetic particles. $I_{X_{i}}$ represents the direct ionization rate while $I_{X_{\alpha}}$ represents the excitation plus ionization rate \cite{Poulin_2017}. Furthermore, the ionization rate $I_X$ can be written in terms of the energy deposition rate from the additional injection of energetic particles due to the aforementioned exotic processes.
\begin{align}
    I_{X_{i}}&=\frac{1}{n_H(z)E_0}\frac{dE}{dV dt}\bigg |_{dep,i}\\
    I_{X_{\alpha}}&=\frac{(1-\mathcal{P})}{n_H(z)E_\alpha}\frac{dE}{dV dt}\bigg |_{dep,\alpha},
\end{align}
where $\mathcal{P}$ is the Peebles coefficient given in the Appendix, $n_H(z)$ is the number density of hydrogen nuclei (proton density), $E_0$ is the ionization energy of a hydrogen atom and $E_\alpha$ is the Lyman-$\alpha$ energy of a hydrogen atom.
Here, we neglect the effect of the extra energy injection on helium ionization, which has been demonstrated to be subdominant and thus should not significantly impact our results.

Using this formalism, we will numerically calculate the evolution of the gas temperature and ionization fraction by incorporating the heating effect induced by the radiative scattering of UHE neutrinos with the cosmic neutrino background.
For solving the differential equations, we assume the initial conditions $T_k(z=10000)=T_{\rm CMB}(z=10000)$ and $x_e(z=10000)=1$.  The assumption is justified because, at high redshift, the gas temperature is strongly coupled to CMB temperature, and the gas is fully ionized. 

\section{Energy Injection Rate due to self-scattering of UHE neutrinos}
\label{sec:Energy Deposition Rate due to self-scattering of UHE neutrinos}
As stated in Sec. \ref{sec:Self-interacting UHE neutrinos from decay of heavy DM}, the UHE neutrinos formed from the decay of superheavy DM can interact with the relic cosmic neutrino background present in the Universe. This scattering can also lead to the production of photons at the one-loop level, which can heat the intergalactic gas and alter both the gas temperature $T_k$ and the brightness temperature $T_{21}$. In this section, we will calculate the energy injection rate resulting from the emission of photons during the scattering of UHE neutrinos with C$\nu$B neutrinos.

The C$\nu$B neutrinos are thermally distributed in the Universe with present-day neutrino background temperature of $T_{\nu,0}=1.9$~K and number density per flavor of $n_{\nu_i,0}=112\ {\rm cm}^{-3}$~\cite{Sarkar_1998}. The velocity-averaged cross section of the incident UHE neutrino having energy $E_{\nu_h}$ with C$\nu$B neutrinos can be expressed in the form~\cite{doring2023astrophysical},
\begin{align}\label{thermallyaveraged1}
    \langle\sigma v\rangle=\frac{1}{n_{\nu_i}}\int \frac{d^3p}{(2\pi)^3}f(\vec{p})v_{M\o l}\sigma(s(E_{\nu_h},\vec{p}))
\end{align}
where, $f(\vec{p})$ is the C$\nu$B neutrino momentum distribution and $v_{M\o l}$ is called M$\o$ller velocity. As C$\nu$B neutrinos will have $m_{\nu_i}\gg T_\nu$ in the relevant range of redshift, the center of mass energy would be independent of momentum $\vec{p}$ of C$\nu$B neutrinos. Hence, we can approximate $s\approx 2E_{\nu_h}m_{\nu}$ and $v_{M\o l}=1$. Using this, the integral becomes
\begin{align}\label{thermallyaveraged2}
    \langle\sigma v\rangle=\sigma(2E_{\nu_h}m_{\nu_i}),
\end{align}
where $m_{\nu_i} \sim 0.1$ eV is the mass of active neutrino. Following the procedure from \cite{Araya_2014}, we find the evolution of the number density of the UHE neutrinos ($n_{\nu_h}$) while scattering with C$\nu$B neutrinos as
\begin{equation}\label{sdn_dt}
    \frac{dn_{\nu_h}}{dt} =  n_{\nu_h} n_{\nu_i}\langle\sigma v\rangle,
\end{equation}
where $n_{\nu_i}$ is the number density of a particular flavor of C$\nu$B neutrinos. Considering that a fraction of DM ($f_{\rm DM}$) in the Universe consists of UHE neutrinos resulting from the decay of superheavy DM, the present-day number density of UHE neutrinos can be calculated from
\begin{equation}\label{sdn_dt2}
n_{\nu_h, \rm 0} =  \frac{f_{\rm DM}\Omega_{\rm DM}\rho_c}{m_{\rm DM}},
\end{equation}
where $\Omega_{\rm DM}$ is the present day relic abundance of DM, $\rho_c$ is the critical density of the Universe and $m_{\rm DM}$ is the mass of DM. Since the neutrinos are nonrelativistic,  their number density varies as $n_{\nu_h}=n_{\nu_{h,0}}(1+z)^3$ and $n_{\nu_i}=n_{\nu_{i,0}}(1+z)^3$, where $n_{\nu_{h,0}}$ and $n_{\nu_{i,0}}$ are present-day neutrino density of UHE neutrinos emitted from decay of DM and present-day neutrino density of single generation of C$\nu$B background, respectively. Using this and Eqs.~(\ref{sdn_dt}) and (\ref{sdn_dt2}), we obtain
\begin{equation}\label{eqs-semi-final}
   \frac{dn_{\nu_h}}{dt} = \frac{(1+z)^6f_{\rm DM}\Omega_{\rm DM}\rho_c\langle\sigma v\rangle n_{\nu_{i,0}}}{m_{\rm DM}}
\end{equation}
As almost the entire rest mass of DM is available as part of the energy of neutrinos, the energy injection rate into the IGM will be then given by multiplying Eq.~(\ref{eqs-semi-final}) with $m_{\rm DM}c^2$ for a simple case of DM two-body decay into a pair of UHE neutrinos.
With this, the energy injection rate due to the given process is finally given by
\begin{equation}\label{eqs-final}
    \frac{dE}{dVdt}\bigg|_{inj}=  (1+z)^6f_{\rm DM} \Omega_{DM}n_{\nu_{i,0}}\rho_c c^2\langle\sigma v\rangle
\end{equation}
Utilizing the expression above for the energy injection rate, we can assess the impact of heating induced by the radiative scattering of UHE neutrinos into photons on the 21-cm brightness temperature by following the general procedure outlined in the previous section.
We assume $f_{\rm DM}\approx 1$ throughout the text, unless explicitly stated otherwise.

\section{Results}
\label{sec:Results and Discussion}
In this section, we present our numerical findings on the evolution of the 21-cm brightness temperature in the context of radiative scattering of UHE self-interacting neutrinos into photons. Given that the brightness temperature is influenced by the evolution of gas temperature and the free electron fraction, we begin by discussing the evolution of these parameters in the presence of energy injection resulting from the radiative scattering.

\begin{figure}[htbp]
    \centering
    \includegraphics[width=0.5\textwidth]{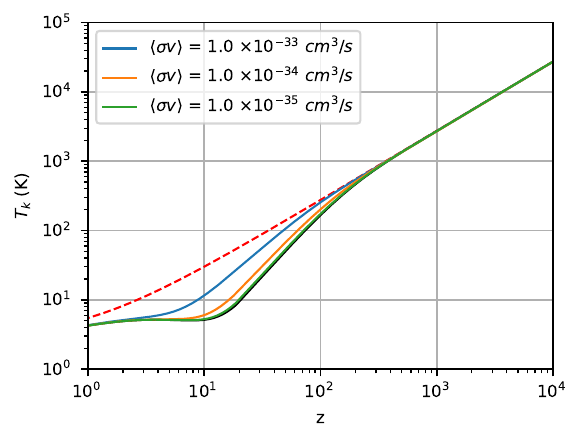}
    \caption{The red dashed line shows the evolution of CMB temperature as a function of redshift ($z$). The black solid curve shows the evolution of neutral hydrogen gas temperature ($T_k$) as a function of $z$ in standard cosmology. The green, orange, and blue curves represent the evolution of $T_k$ for specific reference values of the increasing scattering cross section $\langle \sigma v \rangle$, respectively. The behavior of $T_k$ indicates that increasing the scattering cross section results in the heating of the gas at lower redshifts.}
    \label{fig-Tk}
\end{figure}

By inserting the expression of energy injection from Eqs.~(\ref{eqs-final}) through (\ref{energydeposited}) in Eqs.~(\ref{odeTk}) and (\ref{odexe}), we determine the evolution of gas temperature in standard cosmology (absence of additional heating) as well as in the presence of heating effect induced due to scattering of UHE neutrinos by considering the certain specific value of scattering cross section and $f_{\rm DM} \approx 1$. The solid black curve in Fig.~(\ref{fig-Tk}) represents the evolution of gas temperature while the dashed red line shows the evolution of CMB temperature with redshift in the standard cosmology. The behavior of gas temperature in standard cosmology can be understood as follows: at higher redshift, the Universe was hot, the hydrogen gas was fully ionized and there was no neutral hydrogen. The gas was in equilibrium with the CMB. Thus, the black solid curve coincides with the dashed red line. As the Universe expanded and cooled at redshift around $z \approx 150$, the gas became nonrelativistic and started cooling adiabatically as $T_k\propto(1+z)^2$ while CMB photons cooled as $T_{\rm CMB}\propto(1+z)$.  Thus, the gas temperature starts decreasing faster than the CMB temperature.  At lower redshift $z\approx17$, the gas due to the new mechanism of Lyman-$\alpha$ that is described by Eq.~(\ref{odeTk2}), is heated and is depicted by the plateau region of the curve. Now, the presence of energy injection due to radiative scattering, described by Eq.~(\ref{energydeposited}) and subsequently Eq.~(\ref{eqs-final}), causes the gas temperature to deviate from the standard cosmological behavior. Specifically, as the scattering cross section $\langle \sigma v \rangle$ exceeds $10^{-35}$ cm$^3$ s$^{-1}$, the gas temperature starts to rise at lower redshifts, as depicted in the  Fig.~\ref{fig-Tk}. The green, yellow, and blue curves represent the rise in gas temperature in redshift roughly between $z \approx 150$ and $z\approx 6$ due to an increase in the cross section from $10^{-35}$$-$$10^{-33}$ cm$^3$ s$^{-1}$, respectively. We have verified that a further significant increase in the cross section would completely eliminate the 21-cm absorption signal.

\begin{figure}[htbp]
    \centering
    \includegraphics[width=0.5\textwidth]{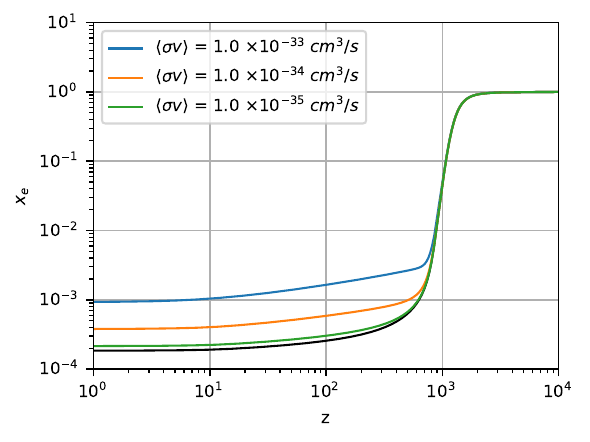}
    \caption{The black solid curve shows the evolution of ionization fraction ($x_e$) as a function of redshift ($z$) in standard cosmology. The green, orange, and blue curves represent the evolution of $x_e$ for specific reference values of the increasing scattering cross section $\langle \sigma v \rangle$, respectively. The behavior of $x_e$ indicates that increasing the scattering cross section results in higher values of $x_e$ at lower redshifts.}
    \label{fig-xe}
\end{figure}

Further, by using Eqs.~(\ref{odeTk}), (\ref{odexe}), and (\ref{eqs-final}) through Eq.~(\ref{energydeposited}), we determine the evolution of the ionization fraction for the given reference values of the scattering cross section and $f_{\rm DM} \approx 1$. The solid black line in Fig.~\ref{fig-xe} represents the evolution of the ionization fraction under standard cosmological conditions. In standard cosmology, at higher redshifts, hydrogen remains ionized, resulting in $x_e = 1$. As the Universe expands and cools, electrons gradually recombine with hydrogen nuclei to form neutral hydrogen atoms. This process reduces the number of ionized hydrogen atoms and free electrons in the Universe. Consequently, the ionization fraction $x_e$ begins to decrease at lower redshifts. In the presence of an additional heating effect due to energy injection from the radiative scattering of UHE neutrinos, the ionization fraction starts to increase at lower redshifts in comparison to the standard cosmology. The green, yellow, and blue curves in Fig.~\ref{fig-xe} represent the rise in ionization fraction at lower redshifts.

\begin{figure}[h]
    \centering
    \includegraphics[width=0.5\textwidth]{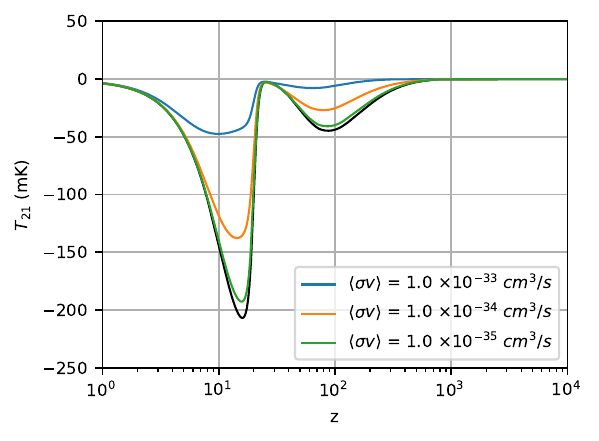}
    \caption{The black solid curve shows the evolution of 21-cm brightness temperature ($T_{21}$) as a function of redshift ($z$) in standard cosmology. The green, orange, and blue curves represent the evolution of $T_{21}$ for specific reference values of the increasing scattering cross section $\langle \sigma v \rangle$, respectively, illustrating the effect of external heating due to the radiative scattering of UHE self-interacting neutrinos. }
    \label{fig-T21}
\end{figure}

Finally, by using Eqs.~(\ref{eq:T21}) and (\ref{spintemperature}), we numerically calculate the 21-cm brightness temperature as a function of redshift. 
The solid black curve in Fig.~\ref{fig-T21} shows the evolution of brightness temperature with redshift in standard cosmology. 
It shows two absorption signals with the first absorption minima near  $z \sim 70$ and the other at higher frequencies at $z \sim 12$$-$$17$ in the global cosmological 21-cm signal.
From Eqs.~(\ref{eq:T21}) and (\ref{spintemperature}), we can see that the evolution of $T_{21}$ depends on the competition between the spin temperature ($T_s$) and CMB temperature. At higher redshifts, the absence of neutral hydrogen prevents spin-flip interactions, thus no absorption or emission of the 21-cm line occurs. Around $z \approx 200$, during the early dark ages, the spin temperature ($T_s$) couples with the gas temperature. As the gas cools more rapidly than the CMB temperature, $T_s$ becomes less than $T_{\rm CMB}$, resulting in a noticeable absorption dip in the 21-cm line at around $z \sim 70$. As the Universe continues to expand and cool, the interaction between the spin temperature and the gas temperature weakens, leading $T_s$ to approach $T_{\rm CMB}$, and no discernible signal is observed. However, at a significantly lower redshift of approximately $z \approx 17$, the Wouthuysen-field mechanism becomes dominant. This mechanism facilitates a strong coupling between the spin temperature and the gas temperature once more. Given the substantial cooling of the gas by this stage, this coupling manifests as a much deeper second absorption signal. 
Finally, this absorption dip concludes as x-ray heating from the first star formation elevates the gas temperature above that of the CMB ($T_s \geq T_{\rm CMB})$. This process could potentially lead to an emission signal, depending on the reionization history driven by the first sources of light in the Universe~\cite{Pritchard_2012}.

\begin{figure}[h]
    \centering
    \includegraphics[width=0.5\textwidth]{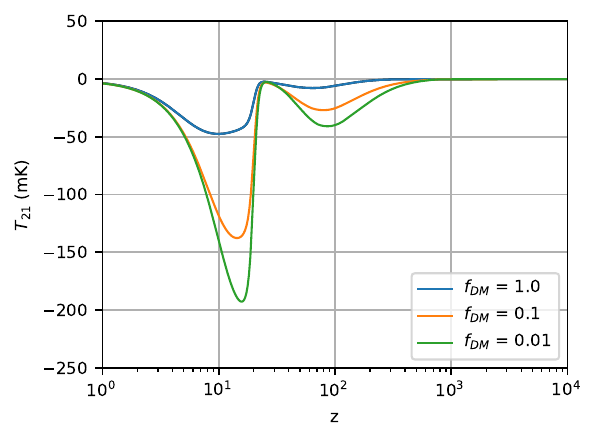}
    \caption{The solid blue, orange and green curves depict the evolution of $T_{21}$ for $f_{\rm DM} = 1, 0.1, 0.01$, with the scattering cross section set to $\langle \sigma v \rangle = 10^{-33}$ cm$^3/s$.}
    \label{fig-T21frac}
\end{figure}

The green, orange, and blue curves in Fig.~\ref{fig-T21} illustrate the effect of external heating due to the radiative scattering of UHE self-interacting neutrinos. When the thermally averaged cross section surpasses a specific reference value, the resulting increase in gas temperature from energy deposited through photon emission can lead to weaker absorption dips in the 21-cm signal. As the scattering cross-section continues to rise, the induced heating can ultimately eliminate the absorption of the 21-cm hydrogen line. 
 We also show the evolution of $T_{21}$ for various values of the dark matter fraction $f_{\rm DM}$ in Fig.~\ref{fig-T21frac}.  Since the energy injection rate in Eq.~(\ref{eqs-final}) is directly proportional to both $f_{\rm DM}$ and $\langle\sigma v\rangle$, reducing $f_{\rm DM}$ has a similar impact on the evolution of gas temperature $T_k$, ionization fraction $x_e$ and the brightness temperature $T_{21}$, as decreasing the scattering cross-section by the same factor. In Fig.~\ref{fig-T21frac}, we can see that decreasing $f_{\rm DM}$ to 0.1 has a comparable effect on the $T_{21}$ as changing cross section  $\langle \sigma v \rangle$ from $10^{-33}$$-$$10^{-34}~{\rm cm^3} s^{-1}$.

Overall, the analysis constrains the scattering cross section value, affecting the brightness temperature magnitude. This suggests that future experiments measuring the 21-cm brightness temperature could offer valuable insights into self-interacting neutrino coupling. By examining the magnitude and characteristics of the 21-cm absorption dips, these experiments can constrain the scattering cross section of UHE self-interacting neutrinos. This, in turn, can provide bounds on the self-interacting coupling strength and the mass of the mediating particle.

 \begin{figure*}[htbp]
  \centering
  \hspace*{-1cm} 
 \subfigure[ $~$ $E_{\nu_h} = 1~{\rm PeV}$]{\includegraphics[width = .5\textwidth,keepaspectratio]{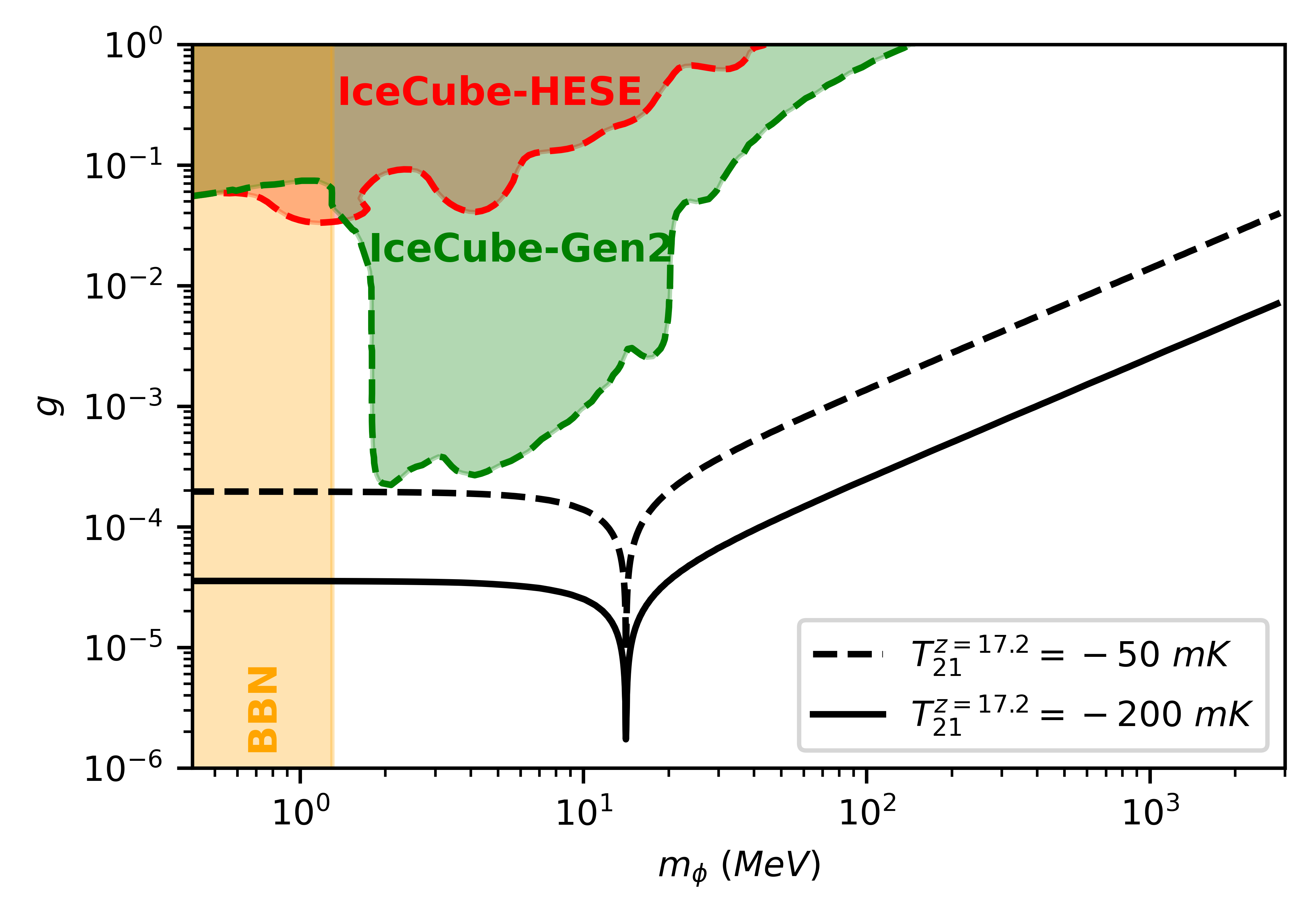}} \quad
  \subfigure[ $~$ $E_{\nu_h} = 10~{\rm PeV}$]{\includegraphics[width = .5\textwidth,keepaspectratio]{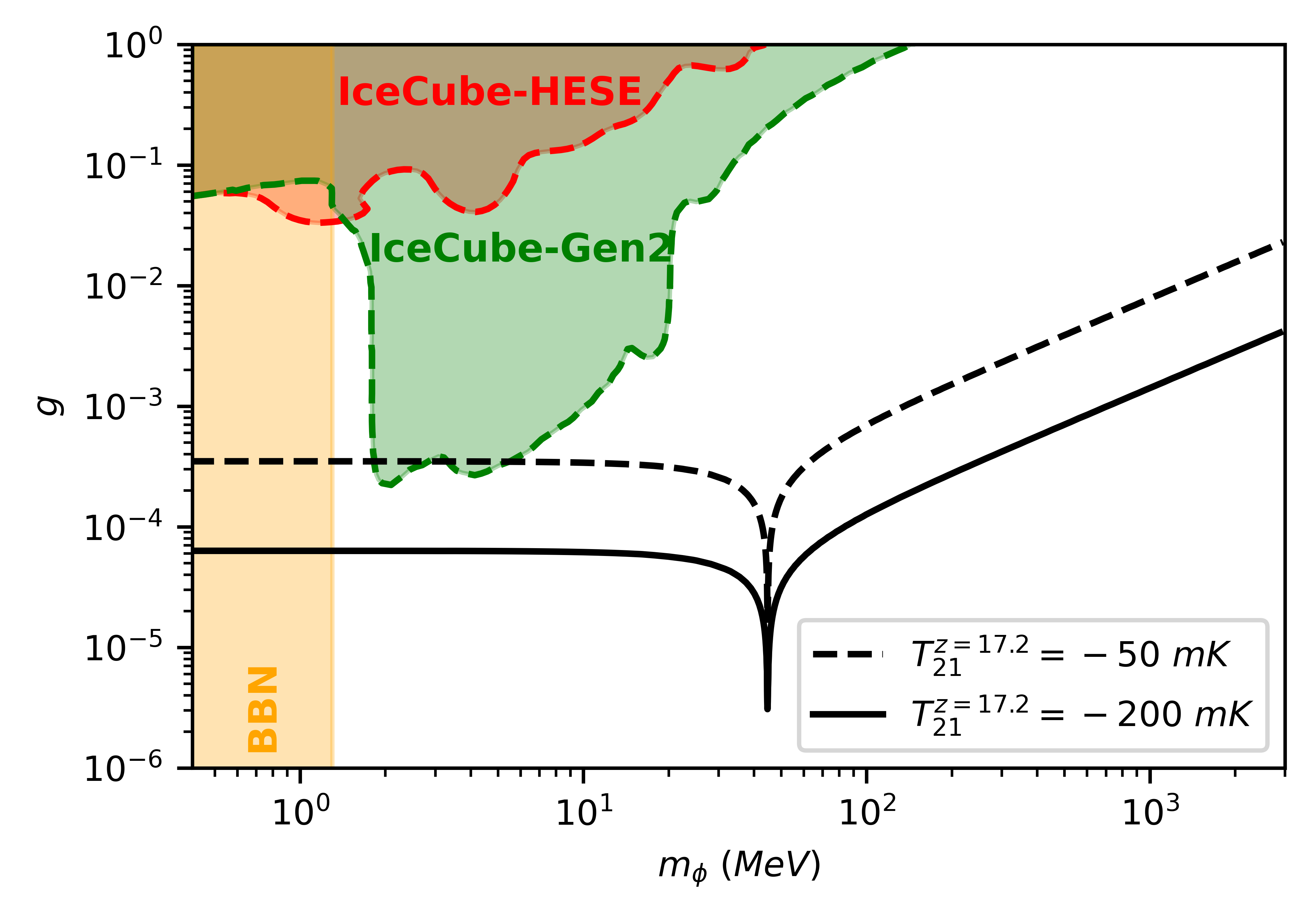}}  
  \hspace*{-1cm} 
  \subfigure[$~$ $E_{\nu_h} = 100~{\rm PeV}$]{\includegraphics[width = .5\textwidth]{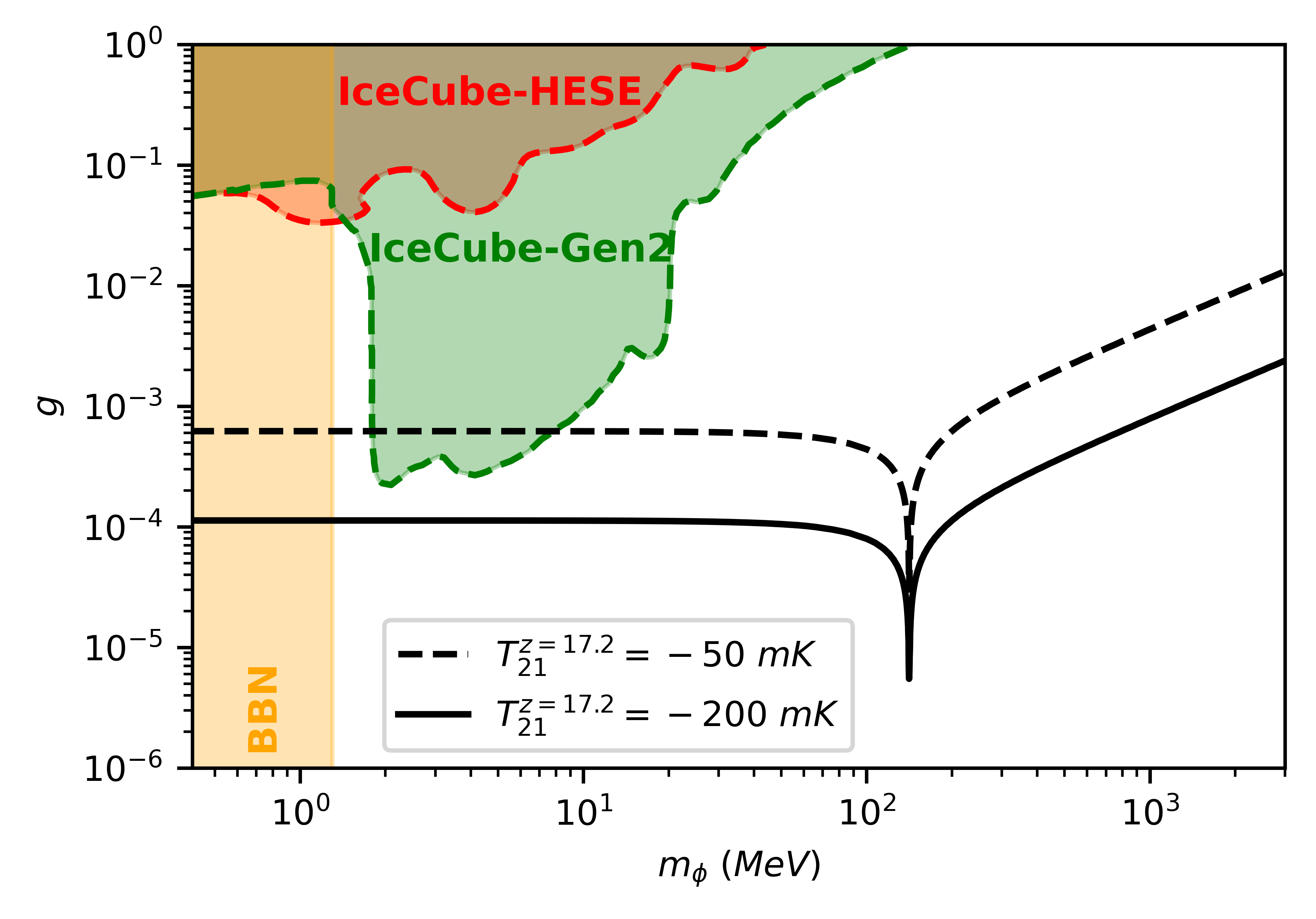}} \quad
   \subfigure[$~$ $E_{\nu_h} = 1~{\rm EeV}$]{\includegraphics[width = .5\textwidth]{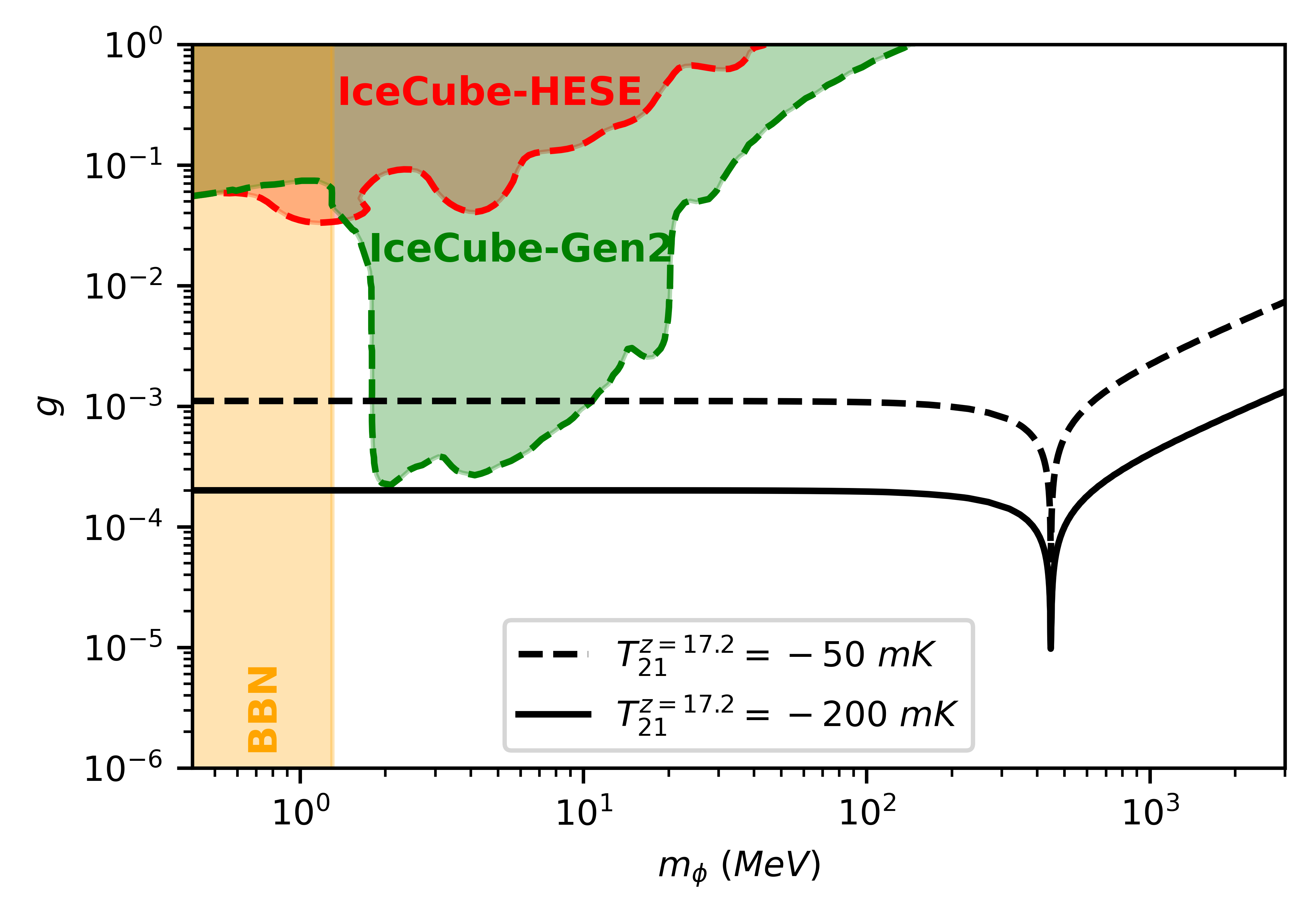}}
  \caption{
Allowed parameter space of self-interacting neutrino coupling ($g$) as a function of the mass of mediator ($m_\phi$) is shown for a fixed value of brightness temperature $T_{21}$ for (a) $E_{\nu_h} = 1~{\rm PeV}$, (b) $E_{\nu_h} = 10~{\rm PeV}$, (c) $E_{\nu_h} = 100~{\rm PeV}$, and (d) $E_{\nu_h} = 1~{ EeV}$, respectively. The red region
bounded by a red dashed line corresponds to the sensitivity for self-interacting $\tau$ neutrinos by using 7.5 years
of the IceCube-HESE data. The green region bounded by a green dashed line is the predicted sensitivity for $\nu_{\tau}$ self-interactions based on simulated data from 10 years of the IceCube-Gen2. The orange region shows the parameter space excluded by constraints from big bang nucleosynthesis.}
   \label{parameterspace}
\end{figure*}

\section{Discussion}
In standard cosmology, the stronger absorption dip at redshift 
$z=17.2$ measures approximately 
$T_{21} \approx -200$ mK. However, as noted earlier, the heating effects from neutrino interactions can diminish the strength of this absorption dip. These interactions introduce additional energy into the intergalactic medium, raising its temperature. Consequently, the contrast between the gas temperature and CMB temperature is reduced, leading to a weaker 21-cm absorption signal.
Here, we use benchmark cross section values required to keep the brightness temperature $T_{21}\approx -200$ mK and $T_{21}\approx -50$ mK at $z=17.2$, respectively. By utilizing these cross section values, we constrain the self-interacting neutrino coupling as a function of the mediator mass. Since there are already quite stringent constraints on self-interacting coupling for muon and electron neutrinos~\cite{Blinov:2019gcj}, in this study, we specifically focus on the self-interacting coupling for $\tau$-generation neutrinos and compare it with the sensitivity of the couplings obtained from 10 years of IceCube data given in~\cite{Esteban_2021}. 
In case of $g_{l_i}$, the experimental constraints exist only on the value of $g_{\mu}$ from the observed $(g-2)_\mu$ measurements. However, these coupling constants typically depend on lepton mass, so $g_{e}$ and $g_{\tau}$ cannot be expected to be same as $g_{\mu}$. For simplicity, we have assume $g_{\nu_\tau} = g$.

According to Eq.~(\ref{scatteringcrosssection}), the scattering cross section also depends on the energy of neutrinos emitted from the decay of the DM candidate. Considering superheavy DM with a mass range between PeV and EeV, the decay of DM would produce neutrinos with energies on the order of PeV to EeV. Consequently, we compute the cross section for $E_{\nu_h} \sim  {\rm PeV}$$-$${\rm EeV}$ and determine the parameter space of self-interacting $\tau$-neutrino coupling as a function of the mediator mass for a specific value of the energy of UHE neutrinos. The results are shown in Fig.~\ref{parameterspace} for the specific value of the energy of UHE neutrinos. The black solid and dashed curves in each subfigure of Fig.~\ref{parameterspace} represent the parameter space of self-interacting neutrino coupling ($g$) and mediator mass ($m_\phi$)
 that satisfy these brightness temperature constraints $T_{21}\approx -200$ mK and $T_{21}\approx -50$ mK at $z=17.2$, respectively. These constraints depend significantly on the energy $E_{\nu_h}$
  of ultrahigh-energy neutrinos generated from the decay of superheavy dark matter. The dip in both curves occurs due to the resonance in the cross section for a particular value of $E_{\nu_h}$ and $m_{\phi}$. 
Interestingly, we notice that before hitting the resonance, for $s > m^2_{\phi}$, the effective cross section $\sigma \propto  {g^4}/{ E_{\nu_h}}$ becomes independent of the mass of mediator. Consequently, the coupling constant $g$ remains nearly constant and increases as $g \propto {E^{\frac{1}{4}}_{\nu_h}}$ for a fixed value of the cross section. After passing through the resonance, where $s < m^2_{\phi}$,  the cross section ${\sigma \propto \frac{ g^4 E_{\nu_h}} {m^4_\phi}}$. Therefore, $g$ increases linearly with the mediator mass $m_{\phi}$
 for a fixed energy value and $f_{\rm DM}$, and the ratio $g/m_{\phi}$ decreases as ${E^{-\frac{1}{4}}_{\nu_h}}$ when the energy of ultrahigh-energy neutrinos $E_{\nu_h}$ is increased. Overall, we observe that across most of the parameter space, as $E_{\nu_h}$ increases from PeV to EeV, the value of $g$ becomes constrained roughly in the range between $10^{-4} - 10^{-3}$ for $f_{\rm DM} \approx 1$. In the most exotic scenarios, if ultrahigh-energy neutrinos $E_{\nu_h}$ reach the grand unification scale, the coupling $g$ could potentially increase up to approximately $0.01$$-$$0.1$, considering the given range of mass for the scalar mediator.

  \begin{figure*}[htbp]
  \centering
  \hspace*{-1cm} 
 \subfigure[$~~E_{\nu_h} = 1~{\rm PeV}, g_{\tau} \neq g_{\nu_{\tau}}, g_{\tau} = 0.1, f_{\rm DM} = 1$]{\includegraphics[width = .5\textwidth,keepaspectratio]{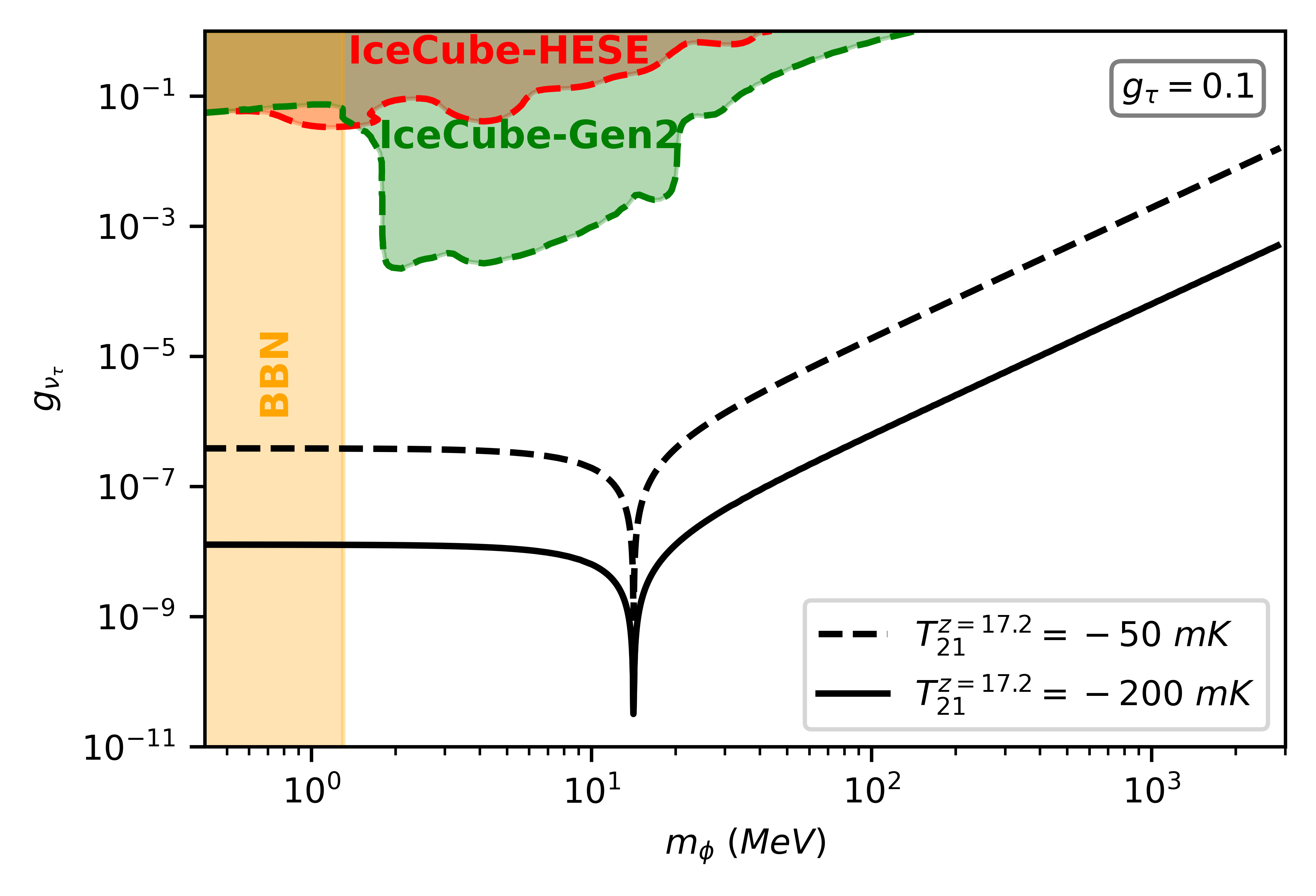}} \quad
  \subfigure[$~~E_{\nu_h} = 1~{\rm PeV}, g_{\tau} = g_{\nu_{\tau}}, f_{\rm DM} = 0.01$]{\includegraphics[width = .5\textwidth,keepaspectratio]{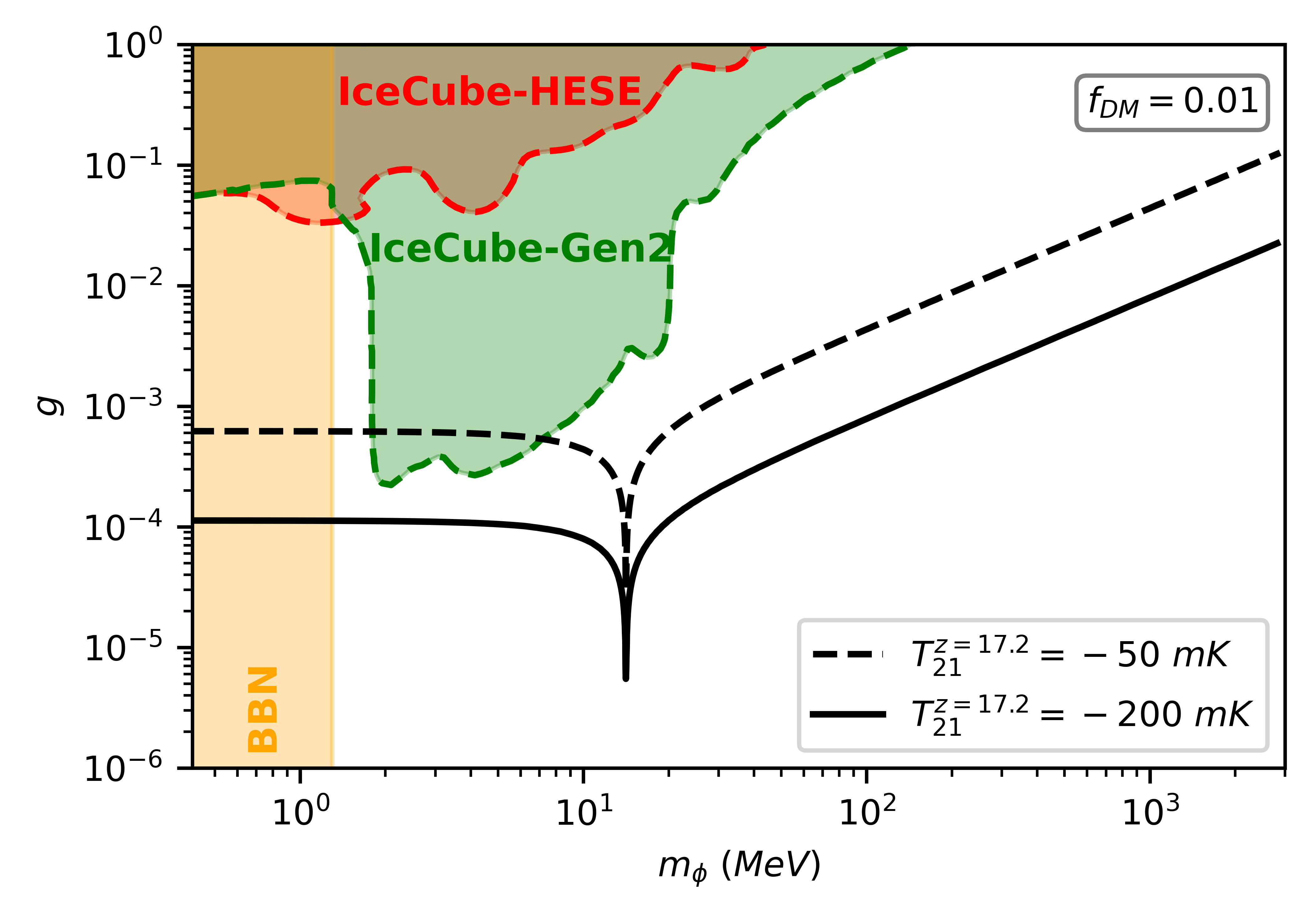}}  
  \caption{Allowed parameter space of self-interacting neutrino coupling (g) as a function of the mass of mediator ($m_\phi$) for the case of (a) $g_{\tau} \neq g_{\nu_{\tau}}$  with $g_{\tau} = 0.1$ and $f_{\rm DM} = 1$, and (b) $f_{\rm DM} = 0.01$ with $g_{\tau} = g_{\nu_{\tau}} = g$, for a fixed energy $E_{\nu_h} =1$ PeV.}
   \label{parameterspace2}
\end{figure*}

 While doing the aforementioned analysis, we have assumed  $g_{\tau} = g_{\nu_{\tau}}$. However, if $g_{\tau} \neq g_{\nu_{\tau}}$,  $g_{\tau}$ could be either larger or smaller than $g_{\nu_\tau}$. Since there are no observational constraints on $g_{\tau}$, one can expect $g_{\tau}$ to be larger than $g_{\nu_\tau}$. As the brightness temperature puts constraints on $(g^4 \equiv g^2_{\nu_\tau} g^2_{\tau})$, increasing $g_{\tau}$ would result in a further reduction in the range of allowed self-interacting $\tau$-neutrino couplings in Fig.~\ref{parameterspace}. Conversely, if theoretical mechanisms suggest a very small 
 $g_{\tau}$, this would increase the allowed range of the coupling parameter $g_{\nu_\tau} = g$.
Additionally, we have assumed $f_{\rm DM} \approx 1$ in this analysis. If the superheavy dark matter constitutes only a small fraction, such as $f_{\rm DM} \sim 0.1 -0.01$, then to maintain a constant brightness temperature of $T_{21}\approx -200$ mK and $T_{21}\approx -50$ mK at $z=17.2$, respectively, a decrease in the value of $f_{\rm DM}$ will need to be compensated by a corresponding increase in the value of $\langle \sigma v \rangle$. Since $\langle \sigma v \rangle \propto g^{1/4}$ for fixed values of $m_{\Phi}$ and $E_{\nu_h}$, decreasing $f_{\rm DM}$ will lead to an increase in $g$ by a factor of $f^{1/4}_{\rm DM}$. As a result, considering $f_{\rm DM} \sim 0.1$$-$$0.01$ will slightly raise the value of g by a factor of approximately ${\cal O}(1)$. For direct comparison of our results, we have plotted the parameter space of $g$ versus $m_{\phi}$ for the case of $g_{\tau} \neq g_{\nu_{\tau}}$  with $g_{\tau} = 0.1$ and $f_{\rm DM} = 1$ in Fig.~\ref{parameterspace2}(a) and $f_{\rm DM} = 0.01$ with $g_{\tau} = g_{\nu_{\tau}} = g$ in  Fig.~\ref{parameterspace2}(b), for a fixed energy $E_{\nu_h} =1$ PeV.

In Figs.~\ref{parameterspace} and \ref{parameterspace2}, we also present the bounds on $g$ from astrophysical and cosmological observations. As mentioned in previous sections, the scattering of UHE neutrinos and C$\nu$B neutrinos results in distinctive dips and bumps in the astrophysical spectrum. Comparing this spectrum with current data from IceCube provides bounds on the self-interacting coupling~\cite{Esteban_2021}. The red region bounded by a red dashed line corresponds to the sensitivity for self-interacting $\tau$ neutrinos by using 7.5 years of the IceCube-HESE data. The green region bounded by a green dashed line is the predicted sensitivity for $\nu_\tau$ self-interactions by considering 10 years of the IceCube-Gen2 (2$\sigma$)~\cite{Esteban_2021}. This indicates that Gen2 of IceCube will be much more sensitive to larger parameter space than contemporary experiments. Further, the interaction between neutrinos and a scalar mediator allows the mediator to be in thermal equilibrium before neutrino decoupling, thereby affecting the relativistic degrees of freedom ($\Delta N_{\rm eff} \lesssim 0.5$) in the Universe. The constraint sets a lower bound on the mediator mass of $m_{\phi} \geq 1.6$ MeV~\cite{Blinov:2019gcj}. The excluded region is depicted as a light orange-shaded band in Fig.~\ref{parameterspace}. There have been constraints from colliders for the $\tau$ generation of neutrinos, but these constraints are much weaker compared to the sensitivity of coupling given by IceCube. Therefore, we have not included them in the Fig.~\ref{parameterspace}. Our results show that the allowed range of self-interacting coupling $g$ is more severely constrained from the detection of absorption brightness temperature $T_{21}$ than the existing IceCube constraints.  

This indicates that the epoch of dark ages and cosmic dawn can potentially provide more competitive bounds compared to existing dominant bounds from other cosmological and astrophysical probes. Another significant advantage of using global brightness temperature measurements is that the dark ages and cosmic dawn periods are relatively free from the complex and often uncertain astrophysical processes that can obscure other signals. This clarity makes this method a promising avenue for investigating UHE neutrino fluxes and potential nonstandard neutrino interactions. 

However, there are certain caveats that need to be considered. In this study, we focus exclusively on neutrinos produced from DM decay. It should be noted, however, that UHE neutrinos can also originate from the decay of primordial black holes (PBHs). In such a case, the relative abundance of PBHs and superheavy DM would affect the constraints on the coupling. Similarly, if superheavy DM can decay or annihilate directly into photons in a specific particle physics model, the impact of these processes would need to be incorporated into the evolution of the 21-cm brightness temperature. As this is beyond the scope of the current work, we leave it for a more detailed investigation in future studies.

\section{Concluding Remarks}\label{sec:Concluding Remarks}
UHE neutrinos, with energies ranging from PeV to EeV, play a crucial role in both cosmology and astrophysics. These particles are among the most energetic and least understood in the Universe, providing valuable insights into various fundamental processes. Interestingly, their interactions with CMB neutrinos en route to Earth can provide unique information about potential self-interactions among neutrinos, which are not well-constrained by current models. In this work, we have investigated the constraints on secret self-interactions of neutrinos emitted through the decay of superheavy DM by studying the impact of their interaction with the cosmic neutrino background on the hydrogen 21-cm signal during the period from the cosmic dark ages to cosmic dawn. Since this period is relatively free from astrophysical uncertainties, it allows for a clearer signal when studying UHE neutrino fluxes and nonstandard neutrino interactions. This makes global brightness temperature measurements a promising avenue for advancing our understanding of neutrino properties.  By examining the magnitude and characteristics of the 21-cm absorption dips, these experiments can constrain the scattering cross section of UHE self-interacting neutrinos, thus providing bounds on the coupling strength and the mass of the scalar mediator.

We have conducted a detailed investigation into the allowed parameter space of self-interacting neutrino coupling as a function of the mediator mass by considering a toy model of a light scalar interacting with neutrinos and the leptonic partners. Utilizing specific cross section values to maintain benchmark brightness temperatures of $T_{21} \approx -200$ mK and $T_{21} \approx -50$ mK at a redshift of $z=17.2$, we have constrained the self-interacting coupling of $\tau$ neutrinos for neutrino energies in the PeV to EeV range, which is characteristic of decays from superheavy dark matter candidates.  Our analysis indicates that as $E_{\nu}$ increases from PeV to EeV, the coupling constant $g$ is constrained roughly within the range of $10^{-4}$ to $10^{-3}$ for $f_{\rm DM} \sim 1$. If we assume that the superheavy DM contributes only a small fraction of the total DM such as $f_{\rm DM} \sim 0.01 - 0.1$, then the coupling constant $g$ will increase by a factor of ${\cal O}(1)$.   Overall, these constraints are much stronger than the predicted sensitivity for $\nu_{\tau}$ self-interactions based on simulated data from 10 years of the IceCube-Gen2. Interestingly, this approach not only offers a novel and competitive method for probing neutrino properties but also utilizes the relatively simple astrophysical conditions during the cosmic dark ages and cosmic dawn to offer a clearer signal for studying nonstandard interactions of neutrinos. Consequently, the potential detection of a robust 21-cm signal by upcoming experiments such as LEDA~\cite{Price_2018}, REACH~\cite{deLeraAcedo:2022kiu, Saxena:2024rhu} etc. can provide critical insights into the nature of dark matter and neutrino physics in future.  In this work, we have focused primarily on analyzing the global brightness temperature within the context of 21-cm cosmology. In future work, we plan to extend our analysis to include power spectra and polarization signatures within the 21-cm signal, as well as explore the potential origins of UHE neutrinos from other exotic processes, such as the decay of PBHs and similar phenomena. Additionally, a comprehensive investigation of self-interacting neutrino coupling will require exploring the origin of such interactions within a consistent model of physics beyond the Standard Model.\footnote{Several aspects of UV-complete models that govern these interactions have been examined in~\cite{Dhuria:2011ye, Dhuria:2012bc, Dhuria:2017ihq}.} Incorporating these additional aspects will enable us to gain deeper insights into the nature of self-interacting neutrinos and their impact on the cosmic environment.

\section*{ACKNOWLEDGEMENTS}
M. D. would like to acknowledge support through the  DST-Inspire Faculty Fellowship of the Department of Science and Technology (DST), Government of India under the Grant Agreement No. IFA18-PH215.

\begin{appendices}

 \section{}
\label{appendix-Ionization-fraction}
The parameters $R_s$ and $I_s$, in Eq.~(\ref{odexe}), are the standard recombination rate (from ionized gas to neutral gas) and standard ionization rate (from neutral gas to ionized gas), respectively, and are given by \cite{Chen_2004}
\begin{align}\label{recombination_rates}
    R_s(z)&=\mathcal{P}\left[\alpha_H x_e^2n_H\right]\\
    I_s(z)&=\mathcal{P}\left[\beta_H(1-x_e)e^{-\frac{h\nu_\alpha}{k_bTk}}\right]
\end{align}
where $\mathcal{P}$ is the Peebles coefficient. It represents the probability that an atom in the first excited state reaches the ground state before being completely ionized. This is given by
\begin{equation}\label{Peeblescoefficient}
    \mathcal{P}=\frac{1+K_H\Lambda_H n_H(1-x_e)}{1+K_H(\Lambda_H+\beta_H)n_H(1-x_e)}
\end{equation}
where $K_H=\pi^2/E_\alpha^3H$ and $\Lambda_H=8.22/{\rm sec}$. $K_H$ accounts for the effect of the expansion of the Universe on the Lyman-$\alpha$ photon, $\Lambda_H$ is the decay rate of the hydrogen atoms from the 2S to 1S level and $E_\alpha$ is the Lyman-$\alpha$ energy of a hydrogen atom.

Here, $\alpha_H$ and $\beta_H$ are the recombination coefficient and photoionization coefficient, respectively, which are given by \cite{Bera_2023},
\begin{align}\label{alpha-beta}
    \alpha_H(T_k)&=F\times10^{-19}\left(\frac{at^b}{1+ct^d}\right)\ m^3s^{-1}\\
    \beta_H(T_k)&=\alpha(T_k)\left(\frac{2\pi m_ek_BT_k}{h_p^2}\right)^{3/2}e^{-E_{2s}/k_BT_k}
\end{align}
where, parameters $F$, $a$, $b$, $c$, $d$, and $t$ are given as
\begin{table}[H]
    \centering
    \begin{tabular}{ccc}
        $F=1.14$, & $a=4.309$, & $b=-0.6166$,\\
        $c=0.6703$, & $d=0.53$, & $t=\frac{T_k}{10^4\ K}$.
    \end{tabular}
\end{table}
 
\end{appendices}

\bibliography{references}
\bibliographystyle{unsrt}

\end{document}